\documentclass[graphicx,amsmath,11pt,aps,pra,twocolumn,reprint,showpacs,showkeys]{revtex4-1}
\usepackage{graphicx}
\usepackage{hyphenat}
\usepackage{color}
\usepackage[normalem]{ulem}%\sout{Hello World}
\usepackage{ifthen}
\newcommand{\fig}[1]{Fig \ref{#1}}
\newcommand{\eq}[1]{eq. \ref{#1}}
\newcommand{\Eq}[1]{Eq. \ref{#1}}
\newcommand{\eqsub}[2]{eqs. \ref{#1}-\ref{#2})}
\newcommand{\linecite}[1]{Ref. \cite{#1}}
\newboolean{showcomment}
\setboolean{showcomment}{false}
\ifshowcomment %
	\newcommand{\pico}[1]{\textcolor{red}{PICO:\textbf{#1}}}
	\newcommand{\add}[1]{\textcolor{red}{#1}}
	\newcommand{\delete}[1]{\textcolor{red}{\sout{#1}}}
\else
  \newcommand{\pico}[1]{}
	\newcommand{\add}[1]{#1}
	\newcommand{\delete}[1]{}
\fi

\begin{document} 

\hyphenation{de-vel-op-ment non-li-n-ear all op-ti-cal nor-ma-l-i-za-tion fac-tor cor-res-p-onds Bloch mo-de en-er-gy con-si-de-red pre-sen-ted}

%\preprint{}

\title{Field renormalization in Photonic Crystal waveguides} %Title of paper



\author{Pierre Colman}
\email[]{pierre.colman@espci.org}
%\thanks{}
\affiliation{DTU Fotonik, {\o}rsted plads, Kgs. Lyngby, Denmark\\
 IEF Institut d'Electronique Fondamentale, Universit\'{e} Paris-Sud, Orsay, France}


\pacs{42.65.Sf,42.65.Tg,42.65.Wi,42.70.Qs,42.25.Bs}
\keywords{Photonic Crystal, Nonlinear optics, Slow-light, NLSE}

\date{\today}

\begin{abstract}
A novel strategy is introduced in order to include variations of the nonlinearity into the nonlinear schr\"{o}dinger Equation. This technique, which relies on renormalization, is in particular well adapted to nanostructured optical systems where the nonlinearity exhibits large variations up to two orders of magnitude larger than in bulk material. We show that it takes into account in a simple and efficient way the specificity of the nonlinearity in nanostructures that is determined by geometrical parameters like the effective mode area and the group index. The renormalization of the nonlinear schr\"{o}dinger Equation is the occasion for physics oriented considerations and unveils the potential of Photonic Crystal waveguides for the study of new nonlinear propagation phenomena.
%This new approach is complementary to the other methods which rely on the successive addition of higher order corrective terms; and so can be use in conjunction with them and provides a more accurate modeling.
\end{abstract}

\pacs{}% insert suggested PACS numbers in braces on next line

\maketitle %\maketitle must follow title, authors, abstract and \pacs


\section{Introduction}
\label{sec:Introduction}
%\cite{Tyrell:JModOp05}=>fdtd
%%%%%%%%%%%%%%
% Specificite PhC ==> very large variation
% 2 technique qui ne donne pas pleinemeent de resultat
% physique differente
% equivalent to a weighted contribution (L-Dos)
%%%%%%%%%%%%%%%%%%%%%%%%%%%%%%%%%%%%%%%%%%%%%%%%%%%%%%%%%

Nanostructured optical systems offer a great plasticity because most of their optical properties can be engineered through the design \cite{Dadap:08,doi:10.1021/nl900371r,Gutman:12,Mori2005,Li:08,OFaolain:10,Ebnali-Heidari:09,Colman:12,Wang:11,Wang:12,VUCK01,Kura:08,Noda:09}. Thus the control by the geometry allows one to get on demand properties, precisely adapted to the phenomena one wants to investigate \cite{Yin:07OL,Foster2006,Colman2012,Ebnali-Heidari:09}. This is both interesting for applications where it is then possible to optimize all optical functions to a large extend \cite{Yeom_2008} and for fundamental investigations where the experimental conditions could be precisely set to maximize a given effect.\\ This plasticity comes usually along with large variations of the optical properties over a small bandwidth because effects of geometry depend highly on the actual wavelength size. For instance in Photonic Crystal waveguides, variations of the nonlinear effective Kerr effect up to 100\% can occur over a 10nm bandwidth \cite{colman:PRA14}. Consequently the modeling of the nonlinear propagation of optical pulses in such structures is essential for understanding precisely the interplay between the different effects taking place; such a task is challenging \cite{PhysRevE.64.056604,6123170,Santagiustina:10}.\\ The use of the generalized Nonlinear Schrodinger Equation (GNLSE) is usually preferred to direct nonlinear FDTD simulation, though that latter being more accurate \cite{LiSa:11,Taflove}, because it has very low computation burden and also provides a direct link between the effective coefficient in the GNLSE and the phenomena that are observed. Namely it is straightforward to add phenomena specifically found in semiconductor optics, like for instance the effect of nonlinear absorption and free carriers \delete{; and to assess their impact on the optical system}\cite{Li:11,Baron:09,Husko:09,Monat2009}. Moreover discussion and interpretation are made easier as the coefficients in the GNLSE are derived into effective lengths like the dispersion length $L_d=T_0^2 / \beta_2$, the nonlinear length $L_{NL}=1/(\gamma P_0)$ or the shock formation length $L_{shock}=0.43{L_{NL}}/{|\tau_{NL}|}$ \cite{AndersonPRA82}, etc. The effective lengths \cite{PhysRevA.57.4791} provide a rapid overview on the relative strength of the different competing effects. However, because the GNLSE relies on several approximations, it is important to check that the physics governing the optical system is still correctly described.
\par Higher order nonlinear effects, for weak perturbations, are taken into account by adding successive corrective terms to the initial NLSE. Here we focus on the inclusion of variations of the nonlinear Kerr response with the angular frequency. Using the transformation $\Delta\omega\rightarrow\imath\partial_t$\cite{agrawalNLoptics}, the Taylor expansion of the Kerr coefficient (hence its variation in the frequency domain) can be included into the GNLSE (time domain). Such decomposition -limited to the first order- takes into account self-steepening effects and can be used to model the formation of optical shock front \cite{deOliveira:92,PhysRevA.57.4791}.
\par Despite being easy to implement, such technique suffers from two major limitations.\\ First it uses the derivation of an aggregated effective parameter computed for a given frequency, whereas by essence nonlinearity involves also the interactions between several waves at different frequencies. Notably, not only the effective mode area, but also the variations of the modes overlap -directly related to cross phase modulation (XPM)- are important in integrated optics. A simple Taylor expansion of the aggregated Kerr coefficient cannot take into account that latter effect. This issue was addressed in \linecite{Tran:09} where it was demonstrated that the physics related to variations of the mode area overlap could be in principle retrieved by the addition of extra operators; and that such corrections could actually have a noticeable impact on pulse propagation.\\The second limitation is due to the nature of the Taylor expansion itself: it is intended for weakly varying functions and convergence tends to be slow as soon as the functions exhibit large or non-monotonic variations. In case of large variations of the nonlinear effective coefficients, the best option would then be to split the total bandwidth into sub-domains wherein variations of the nonlinear properties are negligible \cite{MultipleFWMeqs_PRA08,Santagiustina:10}. Thus each sub-domain is described with good accuracy using a GNLSE; then the different equations are connected through ad-hoc parameters that describe with precision effects like cross phase modulation (XPM) or four wave mixing (FWM). This method is especially well suited to describe FWM because for most practical application pumps, signal and idler spectra are well separated \add{; moreover the small number of beams involved allow setting up a set of coupled mode equations with only a few equations}. However, if the spectrum is continuous over a large bandwidth, splitting the initial simulation domain into several pieces is complex because there is no obvious choice for the separation frequency between the sub-domains; and numerical artefact might appear at the junction between two sub-domains.
\par Consequently neither of the two methods gives satisfactory results in case the single pulse bandwidth extends over large variations of the nonlinear effective parameters. This is especially true for dispersion-engineered Photonic Crystal waveguides (PhCWGs) \cite{colman:PRA14} where the nonlinearity arises mainly from the slow light enhancement\cite{PhysRevE.64.056604}. Indeed any changes of the group index is immediately reported through the slow light factor $S^2=(n_g/n_0)^2$\cite{Krauss:07} on the Kerr nonlinearity, which is expressed as $\gamma_{eff}={\omega}n_{2I}/(cA_{eff})*S^2$. Variations of the modal area are also important in PhCWGs, but still much weaker than those associated with the slow light variation. Note that while S ranges from 2 to 10 in PhCWGs \cite{Li:11}, its effect is less sensitive in Nanowires on silicon which in contrast are much more impacted by variations of the modal area. Indeed the slow light effect in Nanowires is roughly $S\approx 1.2$ \cite{Ding:10} and is in fact already implicitly taken into account in the expression of the effective area\cite{Koos:07}.
\par The method we present here consists in defining and solving the GNLSE for another field than the power flux (cf. \fig{Model}). The point is to find the pseudo field that leads to the GNLSE exhibiting minimal variation of the nonlinear coefficient. Especially this technique is well-suited for nanostructured optical systems where the nonlinearity is defined by the effective area and the group index. \add{Previous studies focused on the ab initio definition of the effective parameters of the propagating equation once the eigen mode’s profile is known. It was demonstrated that the presence of slow light must be taken into account in the normalization in order to guarantee that the correct energy flow through the medium is preserved despite the presence of strong material dispersion or slow light \cite{Sipe:IOP09}. This first normalization step is crucial. However, the resulting nonlinear propagation equation will contain also a few pre-factors; this aspect is not important for few modes problems like FWM or parametric generation where one can define a set of coupled equations, but becomes problematic for continuous mode problems. For example the propagation equation can be turned back into the form of a classical NLSE, but at the cost of extra approximations which ultimately alter the accuracy of the numerical solution.  The second step consists then in the renormalization of the energy flow into a variable whose nonlinear propagating equation has a simpler form. Our finding is that instead of computing the propagation of the power flux, it is better to weight it by the photonic density of states. The group index plays an important role in this second normalization step because the first derivative of the band diagram} corresponds at the same time to the group velocity and to the optical density of states.\\First, we will first review how the slow light effect is introduced into the nonlinear propagation equations; and thus insist on the two contributions of slow light on the nonlinear enhancement. We will then introduce our renormalization technique and apply it to the computation of nonlinear pulse propagation in dispersion-engineered PhCWGs.

\section{NLSE}
\label{sec:NLSE}
We start from the Maxwell's equations. As we are dealing with unidirectional guided propagation (i.e. waveguides), we have split the spatial coordinates into transverse coordinates $\vec{r_{\bot}}$ and  $z$ along the waveguide direction.

\begin{align}
		\nabla\times\vec{E}(\vec{r_{\bot}},z,t)&=-\mu_0 ~\partial_t \vec{H}(\vec{r_{\bot}},z,t)  \label{MaxAmp} \\
		\nabla\times\vec{H}(\vec{r_{\bot}},z,t)&=\epsilon(\vec{r_{\bot}},z) ~\partial_t \vec{E}(\vec{r_{\bot}},z,t) + \partial_t \vec{P_{NL}}(\vec{E}) \label{MaxFar}
\end{align}
	
When $\vec{P_{NL}}=0$, the above equations accept a set of eigen solutions in the frequency domain:

\begin{align}
		\vec{e}(\vec{r_{\perp}},z,\omega)&=\vec{e}_{bloch}(r_{\perp},z-n.a,\omega)e^{-\imath\omega t + \imath K(\omega)z}&\hfill\label{SH:e}\\
		\vec{h}(\vec{r_{\bot}},z,\omega)&=\vec{h}_{bloch}(r_{\bot},z-n.a,\omega)e^{-\imath\omega t + \imath K(\omega)z}&\hfill	\label{SH:h}
\end{align}

with $n \in Z$ and \textbf{a} being the PhC lattice parameters. $K(\omega)$ corresponds to the propagation constant at the frequency $\omega$. For periodic waveguides like PhCWGs, the electric and magnetic field distribution is determined by a Bloch mode which is a periodic function with periodicity \textbf{a}. Hereafter, the dependence on $(\vec{r_{\perp}},z)$ of the eigen and Bloch modes is implicitly assumed and we will use the short notation $\vec{e}(\omega)$, $\vec{h}(\omega)$, $\vec{e}_{Bloch}(\omega)$, $\vec{h}_{Bloch}(\omega)$.\\ Similarly as detailed in \cite{MoloneyPRE02}, the forward propagating equation is obtained by multiplying \eqsub{MaxAmp}{MaxFar} by the solution of the unperturbed ($P_{NL}=0$) system (i.e. $e(\omega)^*$ and $h(\omega)^*$). Substracting to each other the two new equations leads, after few algebra -and especially using the identity $\vec{A}.\nabla\times\vec{B} = \vec{\nabla}.(\vec{A} \times \vec{B}) + \vec{B}.\nabla\times \vec{A}$-, to:

\begin{align}
\vec{\nabla}.(\vec{E}\times \vec{h}(\omega)^* - \vec{H}\times \vec{e}(\omega)^*)=&\nonumber \\
-\vec{e}(\omega)\partial_t\vec{P}_{NL}(\vec{E}) - \partial_t (\epsilon_0\epsilon \vec{E}.&\vec{e(\omega)} +\mu_0 \vec{H}.\vec{h(\omega)})
\label{vectorE}
\end{align}

%\specialcell{\text{(foo)}\hfill}
We will now look for a perturbative solution, in the frequency domain, of the form
\begin{widetext}
\begin{align}
\vec{E}(\vec{r_{\bot}},z,\Delta\omega) = A(z,\Delta\omega)e^{-\imath\Delta\omega t}\vec{e}_{Bloch}(\omega)e^{-\imath\omega_{ref} t + \imath K(\omega_{ref})z}
\label{SVEA:e}\\
\vec{H}(\vec{r_{\bot}},z,\Delta\omega) = A(z,\Delta\omega)e^{-\imath\Delta\omega t}\vec{h}_{Bloch}(\omega)e^{-\imath\omega_{ref} t + \imath K(\omega_{ref})z} \label{SVEA:h}
\end{align}
\end{widetext}

Where $\Delta K(\omega)=K(\omega)-K(\omega_{ref})$ and $\Delta\omega=\omega-\omega_{ref}$. Note that the phazor term in \eqsub{SVEA:e}{SVEA:h} is different from the one found in \eqsub{SH:e}{SH:h}, hence we would like to compute directly the evolution of the pulse envelope centered at frequency $\omega_{ref}$.
Also we assume the envelope $\partial_zA/A<<1/a$ varies slowly, so $\partial_zA(z,\omega)$ is constant over a single PhC lattice. We inject \eqsub{SVEA:e}{SVEA:h} into \eq{vectorE} and integrate over a unit PhC cell; this leads to:

\begin{widetext}
\begin{align}
\left(\partial_zA(z,\Delta\omega)-\imath\Delta K(\omega) A(z,\Delta\omega) \right) \iiint_{cell} \vec{z}\cdot(\vec{e}^*_{Bloch}(\omega)\times\vec{h}_{Bloch}(\omega)-\vec{h}^*_{Bloch}(\omega)\times\vec{e}_{Bloch}(\omega)) dr^3 \nonumber \\
= \imath \omega \iiint_{cell} \vec{e}_{Bloch}(\omega)^*\cdot\vec{P}_{NL}(\Delta\omega,\vec{E}) dr^3
\label{NLSE:0}
\end{align}
\end{widetext}

$A(z,\Delta\omega)$ is defined with regard to a central frequency reference $\omega_{ref}$. Thus \eq{NLSE:0} gives directly access to the envelope of the pulse, without the fast oscillations in space and time $\left\{\omega_{ref},K(\omega_{ref})\right\}$. The integral in the left hand side of \eq{NLSE:0} corresponds to twice the integral of the Poynting vector $\vec{\Pi}_\omega$ over the PhC cell. The link between the Poynting vector and the Bloch mode energy is set through the relationship:

\begin{align}
	\iiint_{cell}\vec{z}\cdot\Re(\vec{e}^*_\omega\times\vec{h}_\omega)dr^3&=&\iint_{Surface}& \vec{z}\cdot\vec{\Pi}_{\omega}dr^2*a \nonumber \\
	&=&v_g(\omega)W_\omega&
	\label{Hellman}
\end{align}
$W_\omega$ is a normalization factor, which corresponds to the Bloch mode energy $1/2\iiint_{cell}\epsilon|e_{Bloch}(\omega)|^2+\mu|h_{Bloch}(\omega)|^2dr^3$. 
Finally we have 

\begin{widetext}
\begin{align}
\partial_zA(z,\Delta\omega) = \imath\Delta K(\omega) A(z,\Delta\omega) + \imath \frac{\omega}{2v_g(\omega)W_\omega} \iiint_{cell} \vec{e}_{Bloch}(\omega)*.\vec{P}_{NL}(\Delta\omega,\vec{E})dr^3
\label{NLSE:E}
\end{align}
\end{widetext}

Before going further into details, let us focus an instant on the peculiar normalization choice for both $A(z,\omega)$ and $W_\omega$. \add{In the linear regime, the choice is to preserve the power flux, so the computed field directly corresponds to the energy flowing though the medium. Using \eq{Hellman}, one finds that the total energy transiting through the waveguide at frequency $\omega$ is $P(z,\omega)=|A(z,\omega)|^2v_g(\omega)W_\omega/a$. As a result the Bloch modes are normalized such as $v_g(\omega)W_\omega/a=1$; so one gets directly from the NLSE $P=|A|^2$. However such normalization is done independently of the nonlinear problem which is considered; consequently it may not necessary be the best choice from a pure numerical point of view.}
\par Regarding the nonlinear polarization $P_{NL}$, we assume that it is due to the $\chi^{(3)}$ response expressed as: $\vec{P_{NL}}(r,\omega_0)=3/2\epsilon_0\chi^{(3)}_{111}$$\int_{\omega 1-2-3}(\vec{E}^*(r)_{\omega_1}\cdot\vec{E}(r)_{\omega_2}).\vec{E}(r)_{\omega_3})$$\delta(\omega_2+\omega_3-\omega_1-\omega_0)$ , where $\delta$ stands for the Dirac delta function. The exact form of $P_{NL}$ is directly related to the nonlinear tensor \cite{Boyd_NLO}; and consequently will have a different formulation depending on the material that is actually considered. However the conclusions presented here are general and could be easily extended to any peculiar form of $P_{NL}$. Finally, we obtain

\begin{widetext}
\begin{align}
	\partial_z A(z,\omega_0) =  \imath K(\omega_0) A(\omega_0) + \frac{\imath\omega_0n_{2I}}{cA_{eff}(\omega_{0})}\frac{\sqrt{n_{g\omega_0}}} {n_0^2} \int E(z,\omega_1)^*E(z,\omega_2)E(z,\omega_3)gh_{\omega0}(\omega_1,\omega_2,\omega_3) \delta(\omega_2+\omega_3-\omega_1-\omega_0)d\omega_i^3
	\label{NLSE:Ekhi3}
\end{align}
\end{widetext}

We have defined $n_{2I}=3\chi^{(3)}/(4c\epsilon_0n_0^2)$. We also introduced the error function $gh_\omega$ which takes into account the fact that the mode overlap integrals may deviate from $A_{eff}(\omega_{0})$ for frequency $\omega_i$ different than $\omega_{0}$. Hence 

\begin{align}
	&A_{eff}(\omega_0)= \frac{\left(\iiint_{Cell}\epsilon(r)|e_{\omega0}|^2dr^3\right)^2}{an_0^4\iiint_{NLmat.}|e_{\omega0}|^4dr^3} \label{Aeff} \hfill&\\
	&gh_{\omega0}(\omega_1,\omega_2,\omega_3)= \frac{ \iiint_{NLmat.}(\vec{e}_{\omega0}^*\cdot\vec{e}_{\omega2})(\vec{e}_{\omega1}^*\cdot\vec{e}_{\omega3})dr^3 } {\iiint_{NL mat.}|e_{\omega0}|^4dr^3}&&\label{gh_omega}
\end{align}

$NL\_mat$ indicates that the integral is performed over the nonlinear material in the PhC unit cell. At this point, one notes that the nonlinearity depends on the intensity of  electric field $E$ while in \eq{NLSE:Ekhi3} the nonlinear pulse evolution is set for the power flux $P=|A|^2$ \cite{PhysRevE.64.056604}. \Eq{Hellman} is then used once more and one finally gets:

\begin{widetext}
\begin{align}
	\partial_z A(z,\omega_0) =  \imath K(\omega_0) A(\omega_0) +\imath \frac{\omega_0n_{2I}}{cA_{eff}(\omega0)}\frac{\sqrt{n_{g\omega_0}}} {n_0^2} \int& A(z,\omega_1)^*A(z,\omega_2)A(z,\omega_3)gh_{\omega0}(\omega_1,\omega_2,\omega_3)\nonumber\\
	&\sqrt{n_g(\omega_1)n_g(\omega_2)n_g(\omega_3)} \delta(\omega_2+\omega_3-\omega_1-\omega_0)d\omega_i^3
	\label{NLSE:A}
\end{align}
\end{widetext}

Unfortunately, the $\sqrt{n_g(\omega_{i=1,2,3})}$ and $gh_{\omega0}(\omega_1,\omega_2,\omega_3)$ terms in the right hand side depend on $\omega_{i=1,2,3}$, hence these prefactors cannot be set aside the integral. However, the formulation of a standard NLSE equation -expressed in the frequency domain- can be retrieved assuming that $gh_{\omega}=1$ and that $\sqrt{n_g(\omega_i)}$ do no vary: 

\begin{widetext}
\begin{align}
	\partial_z A(z,\omega_0) =  \imath K(\omega_0) A(z,\omega_0) +\imath \frac{\omega_0n_{2I}}{cA_{eff}(\omega0)}\frac{n_{g\omega_0}^2}{n_0^2} \int & A(z,\omega_1)^*A(z,\omega_2)A(z,\omega_3) \delta(\omega_2+\omega_3-\omega_1-\omega_0)d\omega_i^3	\label{NLSE:omega}
\end{align}
\end{widetext}

Note that \eq{NLSE:omega} could have also been directly obtained through the derivation of \eq{NLSE:A} for a monochromatic wave propagation. In fact this is most of the time how the effective nonlinear coefficient of the NLSE are derived. But the purpose here is precisely to point out explicitly the frequency dependence of the different effective parameters; and the approximations done in regard with it. The propagating equation corresponding to \eq{NLSE:omega} in the time domain is:

\begin{align}
	\partial_zA(z,t) = \imath D(\imath\partial_t)A(z,t) + \imath \gamma_0 |A(z,t)|^2A(z,t)
	\label{NLSE:time}
\end{align}

Here $\gamma_0=\gamma(\omega_{ref})$ is the effective Kerr nonlinearity with  $\gamma(\omega)=n_{2I}\omega/(cA_{eff}(\omega)).(n_g/n_0)^2$. The dispersion operator $D(\imath\partial_t)=\sum_{n\geq 2}(\partial^n_\omega k)(\imath\partial_t)^n/n\!$ accounts for dispersion at all orders, with t being the retarded time in the moving frame at velocity $c/n_{g-\omega_0}$. This NLSE does not include any variations of the effective nonlinearity over the simulation domain. Effects related to the dispersive nonlinearity are then re-introduced using the first order perturbation $\gamma_1$ which is accordingly defined as $\gamma_1=\partial_\omega\gamma(\omega_{ref})$. 

\begin{align}
\partial_zA(z,t) = \imath & D(\imath\partial_t)A(z,t) \nonumber \\
													& + \imath (\gamma_0 + \gamma_1\imath\partial_t) |A(z,t)|^2A(z,t)
	\label{GNLSE}
\end{align}

\Eq{GNLSE} is derived thanks to the fact that the different $\sqrt{n_g(\omega_{i=\{1,2,3\}})}$ factors in \eq{NLSE:A} are set outside the integral sign. Such an operation is only valid if the group index $n_g$ does not depend on the different frequencies $\omega_i$ that contribute to the nonlinear effect. This (mathematical) approximation is equivalent to the hypothesis according which \textbf{the strength of nonlinearity only depends on the angular frequency which the nonlinear effect takes place at; hence not on the different frequencies contributing to the nonlinear effect}. There is no evidence that this assumption is valid in general. Especially when it comes to nanostructures, one may think that the photons that are the most confined contribute the most to the nonlinearity.
Although the equi-contribution hypothesis does not hold in general, it can happen that the $\sqrt{n_g(\omega_1)n_g(\omega_2)n_g(\omega_3)}$ prefactor depends only on $\omega_0$. Indeed the different frequencies $\omega_i$ are not independent, but must satisfy the energy conservation condition $\omega_0+\omega_1=\omega_2+\omega_3$. In such a case the equi-contribution approximation still describes the nonlinear photon dynamics with accuracy; and we will refer to such photons as being \textit{dispersive photons}.

\section{Renormalization}
\label{sec:Renormalization}

\begin{figure}[htb]
\centerline{\includegraphics[width=6cm]{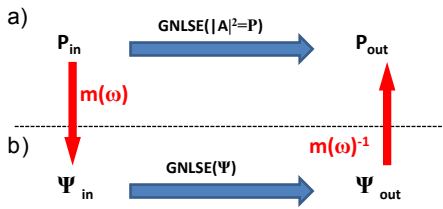}}
\caption{Schematic of the two methods used here to compute the nonlinear pulse propagation in PhC waveguide. a) Standard method: the GNLSE is solved for the energy flux. b) Input conditions (and the GNLSE) are renormed, and the nonlinear propagation is solved for a pseudo-electric $\Psi$ field.}
\label{Model} 
\end{figure}

As depicted by \fig{Model} we shall try to circumvent the problem caused by fluctuations of the nonlinear effective parameters in \eq{NLSE:A} by finding a proper referential wherein the nonlinear response is flat.\add{} Looking back at \eq{NLSE:A} tells us that this is indeed the key in order to obtain a time-domain propagation equation like \eq{NLSE:time} : the different frequencies should have an \textbf{equi-contribution} to the nonlinear process; hence no frequency dependent pre-factor must appear inside the integral sign of \eq{NLSE:A}. To do so, we decided to weight the frequencies by a ad-hoc $m(\omega)$ contribution.\\Note that we neglect at first any variation of the mode field distribution over the pulse bandwidth ($gh_{\omega0}(\omega_i)=1 ~\forall~\omega_0$). The reason for this is that variations of the group index in PhCWGs account for about 75\% of the total variation of the effective nonlinear coefficients \cite{Santagiustina:10,colman:PRA14}. Consequently, dealing with the variations of the slow-light factor would be the first step. Moreover this simplified case is a good test case to present the renormalization technique. We will show in next section how the variations of the mode field distribution can be taken as well into account by this technique.
\par It appears that a natural choice would be to solve the propagation equation for the field $\Psi(\omega)=\sqrt{n_g(\omega)/n_0}A(\omega)$ so \eq{NLSE:A} becomes

\begin{widetext}
\begin{align}
	\partial_z \Psi(z,\omega_0) =  \imath K(\omega_0) \Psi(z,\omega_0)
	+\imath \frac{\omega_0n_{2I}}{cA_{eff}(\omega0)}\frac{n_{g\omega_0}}{n_0} \int& \Psi(z,\omega_1)^*\Psi(z,\omega_2)\Psi(z,\omega_3) \delta(\omega_2+\omega_3-\omega_1-\omega_0)d\omega_i^3
	\label{NLSE:Renormed}
\end{align}
\end{widetext}

%%%%%%%%%%%%%%%%%%%%%%%%%%%%%%%%%%%%%	FIGURE	1 %%%%%%%%%%%%%%%%%%%%%%%%%%%%%%%%
%%%%%%%%%%%%%%%%%%%%%%%%%%%%%%%%%%%%%%%%%%%%%%%%%%%%%%%%%%%%%%%%%%%%%%%%%%%%%%%%%%%

Instead of solving the nonlinear propagation of the power flux $P=|A|^2$, we are now solving the propagation of a pseudo-field $\Psi$. Looking in details at the differences between \eq{NLSE:Renormed} and \eq{NLSE:omega}, we see that the slow light enhancement factor enhancement $S^2=(n_g/n_0)^2$ has been replaced by $S$. Interestingly, if we had introduced higher order nonlinear effects such as three photons absorption (3PA) which are associated to a slow light enhancement of $S^3$, then the slow light pre-factor would have been turned into $S$ as well. A first feature is that the strength of the nonlinearity appears to be much weaker. Besides if we consider the variations of the Kerr effect $\partial_\omega\gamma(\omega)$, the contribution of the slow-light to the characteristic self-steepening time \cite{colman:PRA14} $\tau_{NL}=\partial_\omega\gamma(\omega)/\gamma(\omega)$ has been halved! Consequently \eq{NLSE:Renormed} exhibits weaker nonlinearity and even weaker relative nonlinear dispersion than \eq{NLSE:omega}. Weaker relative variations of the nonlinearity means that the actual nonlinear variations could be taken into account with more accuracy by \eq{NLSE:Renormed}- which could also include second order perturbative corrections-. Note that although the nonlinearity appears much weaker, the input field has been renormalized as well, hence it is much stronger. Consequently global parameters like the soliton number are preserved.
\par Usually, such a strategy is not convenient because it requires additional transformation back and forth between computed quantities ($\Psi\neq A$) and measured ones (power flux $P=|A|^2$) -cf. \fig{Model}. Also one could wonder whether such a renormalization would not simply lead to unphysical solutions, hence simpler equations but not describing correctly the physics. Any renormalization could be applied to \eq{NLSE:A}, as long as it is done in consistency with the math -e.g. the integral sign in \eq{NLSE:A}-; just that not any renormalization would actually lead to a simpler formulation. Regarding the specific choice of $m(\omega)=\sqrt{n(\omega)/n_0}$, \eq{Hellman} tells that because $n_g|A|^2\propto|E|^2$, then the pseudo-field $\Psi$ is actually directly proportional to the Bloch mode electric field. This is consistent with the fact that the electric field density is indeed the physical quantity that matters for nonlinearity, not the power flux. \add{Because the group velocity governs most of the physics (slow-light enhancement) in PhCWGs, it is then not surprising that it plays a role both in the initial normalization of the Bloch field and also here in the renormalization of the computed power flux.} \delete{As a supplementary proof, if we reduce the study to the propagation of four monochromatic beams, hence two pumps, one signal and one idler; \eq{NLSE:Renormed} can then be turned into a set of four coupled equations that corresponds exactly to the one derived in an analytic way in \linecite{Santagiustina:10}, with all the group indices pre-factors taken correctly into account. Thus our technique could also be seen as a generalization of \linecite{Santagiustina:10} that deal in a seamless way with the generation of multiple signal and idler orders, hence not only limited to a set of four beams.}\\ In brief, the renormalization technique poses the question whether the natural choice that is usually made to compute the evolution of the power flux is right. We think it is not for nanostructured systems.

\subsection{Implementation and comparison}
\label{sec:Experiment}
Nonlinear pulse propagation can now be dealt with in two different ways. On one hand, one can use \eq{GNLSE}, expanding eventually the Taylor series of the nonlinearity beyond the first order in order to get a better match; on the other hand \eq{NLSE:Renormed} will also provide an accurate result, and might be easier to implement. The outcome of both equations should be about the same, given that they are indeed describing the same physical system.\\To investigate the differences between \eq{GNLSE} and \eq{NLSE:Renormed}, we take as a test case the nonlinear pulse propagation experiment performed in conditions similar to those in \linecite{Colman2012}. A $2.3$\textit{ps} Fourier limited pulse is send close to the zero group velocity wavelength (ZDW) of an 1.5mm-long \textit{dispersion-engineered} PhC waveguide. The pulse peak power is 8W and corresponds to a soliton number of $N=2.1$.  For consistency with real systems, the equations have been adapted to include the specificity of optical semiconductor nanostructures, mainly the effect of nonlinear absorption and the presence of free carriers \cite{Li:11,Baron:09,Husko:09,Monat2009}. The power flux propagation is computed by mean of the following generalized Schr\"{o}dinger equation\cite{NatPhot_COlman10}.

\begin{align}
\frac{\partial A}{\partial z} = -\frac{\alpha}{2}  A -  \alpha_3 |A|^4A - D(\imath\partial_t)A \nonumber \\
+\imath (\gamma_0 + G(\imath\partial_t)) |A|^2A-(\sigma+ik_0\delta)N~A
\label{GNLSE_Nocited}
\end{align}

$\alpha_0=2dB/mm$ and $\alpha_3=25/W^2/mm$ \cite{Husko:09} stand for the linear propagation loss and three photons absorption (3PA). The dispersion operator $D(\imath\partial_t)=\sum_{n\geq 2}(\partial^n_\omega k)(\imath\partial_t)^n/n\!$ 
accounts for dispersion at all orders, with t being the retarded time in the moving frame at velocity $c/n_g$ (calculated at the input wavelength). Besides, we introduce the Kerr operator $G(\imath\partial_t)=\sum_{n\geq 1}(\partial^n_\omega \gamma)(\imath\partial_t)^n/n!$ to takes into account the dispersion of the Kerr coefficient with the angular pulsation (higher order shock terms). Such expansions are intended to provide accurate numerical results, though it limits the insight into the physical parameters that govern the pulse propagation. \add{$\sigma$ and $\delta$ account for the free carriers absorption and dispersion respectively. Owing to the fact that we are considering here a high bandgap material like GaInP that exhibits solely three photons (ThPA) and no two photons absorption, the self generated plasma does not impact much the overall dynamics. Practical details related to the way the free carriers effects are computed and added to the NLSE are found in  \linecite{Husko_Nat14,Husko_ScRep13,NatPhot_COlman10}.}

%%%%%%%%%%%%%%%%%%%%%%%%%%%%%%%%%%%%%	FIGURE	2 %%%%%%%%%%%%%%%%%%%%%%%%%%%%%%%%
\begin{figure}[htb]
\centerline{\includegraphics[width=9cm]{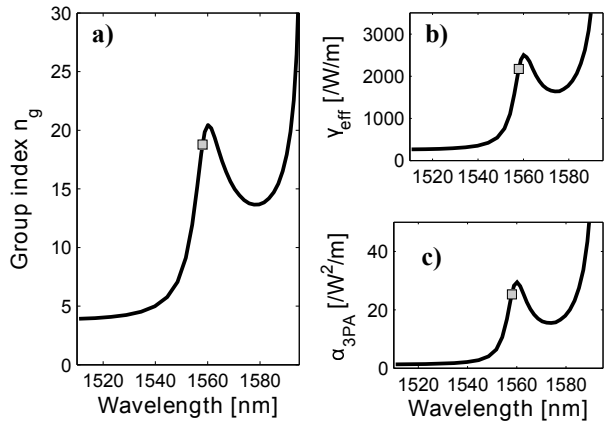}}
\caption{Parameters used for the simulation. a) group index $n_g$ as function of the wavelength. b) Effective Kerr coefficient. c) Effective three photons absorption (3PA) coefficient. The cyan square indicates the position of the input wavelength in \fig{Simulation}}.
\label{Parameter} 
\end{figure}
%%%%%%%%%%%%%%%%%%%%%%%%%%%%%%%%%%%%%%%%%%%%%%%%%%%%%%%%%%%%%%%%%%%%%%%%%%%%%%%%%%%

As a guideline we have $\beta_2=-6.7ps^2/mm$, $\beta_3=-1.7ps^3/mm$, $\gamma_0=2200/W/m$ and $\gamma_1/\gamma_0=-170fs$, the waveguide dispersion and the variation of the nonlinear coefficients (Kerr and ThPA) with the angular frequency are shown in \fig{Parameter}.
%%%%%%%%%%%%%%%%%%%%%%%%%%%%%%%%%%%%%	FIGURE	3 %%%%%%%%%%%%%%%%%%%%%%%%%%%%%%%%
\begin{figure}[htb]
\centerline{\includegraphics[width=9cm]{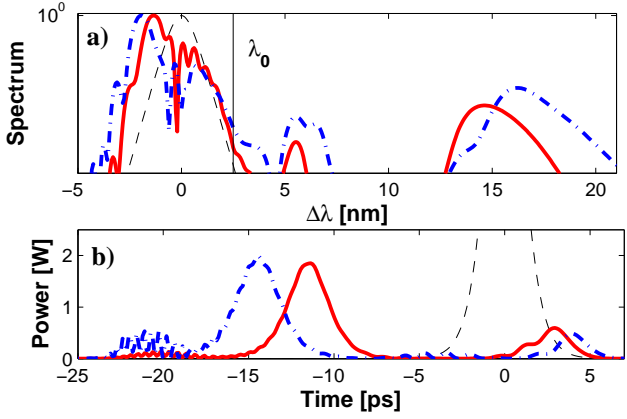}}
\caption{Comparison between the two numerical models. a) Spectra. b) Temporal trace. dashed-black: input pulse; blue: with renormalization $m=\sqrt{n_g(\omega)/n_0}$. red: without renormalization. $\lambda_0$ indicates the position of the zero group velocity dispersion point (ZVD).}
\label{Simulation} 
\end{figure}
%%%%%%%%%%%%%%%%%%%%%%%%%%%%%%%%%%%%%%%%%%%%%%%%%%%%%%%%%%%%%%%%%%%%%%%%%%%%%%%%%%%

Briefly, the renormed NLSE equation is obtained from \eq{GNLSE} by (i) expressing it in the frequency domain, (ii) dividing the nonlinear coefficient by $m(\omega)^n$ where $n=2$ for $\chi^{(3)}$ effect (e.g. Kerr) and $n=4$ for $\chi^{(5)}$ (e.g. Three photons absorption), (iii) the input field $\Psi$ is then obtained by multiplying the initial input field $A(\omega)$ by $m(\omega)$.

\par Comparison between the results of \eq{GNLSE} (red) and its counterpart solved for the pseudo electric field $\Psi$ (blue) are shown in \fig{Simulation}. The input soliton has undergone a blue Soliton-Self Frequency Shift (SSFS) of a few \textit{nm} and dispersive waves are generated in the normal dispersion region \cite{Biancalana:04,PhysRevLett.107.203902}. In the temporal domain the complex interplay \cite{colman:PRA14} of the dispersive nonlinearity \cite{Raineri_PRB13}, free carriers effects \cite{Husko_ScRep13}, and the SSFS, results in a pulse advance of  a few picoseconds. We see the spectral position of the dispersive wave (in the normal dispersion region) and the time advance (10ps vs. 15ps) differ between the two models. Indeed after the initial generation, the dispersive wave interacts with the soliton through cross-phase modulation and cascaded four wave mixing \cite{PhysRevLett.109.223904}: therefore its amplitude and position depends greatly on the exact form of the Kerr nonlinearity. The amplitude of the dispersive waves grows much stronger when the renormalization procedure is employed; and because the spectral recoil appears in reaction to the emission of the dispersive wave, the SSFS is much stronger for the renormalized case. Divergences between the two numerical models are clearly visible.

\subsection{Comparison with an analytic solution}
\label{sec:Analytic}
Now that we have shown that the two approaches (the nominal GNLSE and its renormalized counterpart) lead to different results, we must determine which model is (the most) correct. The case that we just discussed corresponds to a realistic case where both the medium (dispersion, nonlinearity, absorption) and the input parameters (power, duration) are within reach of current experiments. So it would be in principle possible to perform such an experiment and compare the two models with the experimental results. However such measurements are not available yet. Another way would be to confront directly the results of the two models with an analytic case. 
\par Recent work by \textit{Erkintalo et al.} \cite{PhysRevLett.109.223904} demonstrated that nonlinear pulse propagation (continuous spectrum) and cascaded FWM (discrete spectrum) are closely related and that the nonlinear pulse dynamics can be described as a cascade of FWM events. This means that the capacity of an equation to accurately reproduce the reality is intrinsic to its capacity to deal correctly with FWM. Although no analytic solution exists for the exact case we just studied we can still simplify our problem to the propagation in a lossless and $L=100\mu m$ short PhCWG of a $P_0=8W$ single continuous wave beam of same central frequency ($\omega_{ref}$) as previously. In such a situation the short length of the waveguide and the absence of propagation loss render the undepleted pump approximation valid. The FWM conversion efficiency \cite{Ebnali-Heidari:09} depending on the pump signal detuning ($\delta\omega$) is then expressed through the analytic formulation:

\begin{align}
\eta(\delta\omega) =  (\gamma_{FWM}(\delta\omega)P_0L)^2 (&\frac{sinh(g(\delta\omega )L)}{g(\delta\omega )L})^2  &\\
g^2(\delta\omega)  =   (\gamma_{FWM}(\delta\omega)P_0)^2 ~~&& \nonumber \\
- ( \Delta K_{L\omega Ref}(\delta\omega) + &\Delta K_{NL\omega Ref}(\delta\omega))^2/4 & \\
\Delta K_{L\omega Ref}(\delta\omega)  =  2K(\omega_{ref})-K&(\omega_{ref}+\delta\omega) & \nonumber \\
&-K(\omega_{ref}-\delta\omega) & \\
\Delta K_{NL\omega Ref}(\delta\omega) = 2P_0(\gamma_{XPM}(\delta&\omega) +\gamma_{XPM}(-\delta\omega) & \nonumber \\
&- \gamma_{SPM} (\omega_{ref})) &
\label{dK_NL}
\end{align}

$\gamma_{FWM}$, $\gamma_{XPM}$, $\gamma_{SPM}$ account respectively for the FWM nonlinear coupling coefficient, the cross phase modulation (XPM) between the strong pump and the weak signal/idler and the self-phase modulation (SPM) of the pump \cite{ Santagiustina:10}.  Especially it takes into account the different modes fields overlap. According to the notation used previously, these effective nonlinear coefficients are expressed as: 

\begin{align}
\gamma_{SPM}(\omega_{ref})&{=}  \frac{\omega_{ref} n_{2I}}{cA_{eff}(\omega_{ref}) } gh_{\omega{ref}}(\omega_{ref},\omega_{ref},\omega_{ref})  \nonumber \\
& (\frac{n_g(\omega_{ref})}{n_0})^2 \\
\gamma_{XPM}(\omega)&{=}   \frac{\omega  n_{2I}}{cA_{eff}(\omega) } gh_{\omega{ref}}(\omega_{ref},\omega,\omega_{ref})  \nonumber    \\
& (\frac{n_g(\omega_{ref}) n_g(\omega)}{n_0^2}) \\
\gamma_{FWM}(\delta\omega)&{=}   \frac{(\omega_{ref}{+}\delta\omega) n_{2I}}{cA_{eff}(\omega{+}\delta\omega) } gh_{\omega_{ref}}(\omega_{ref}{-}\delta\omega,\omega_{ref},\omega_{ref}) \nonumber \\
& (\frac{ n_g(\omega_{ref}) \sqrt{ n_g(\omega_{ref}{-}\delta\omega) n_g(\omega_{ref}{+}\delta\omega)}  }{n_0^2})
\label{gamma_FWM}
\end{align}

\begin{figure}[htb]
\centerline{\includegraphics[width=9cm]{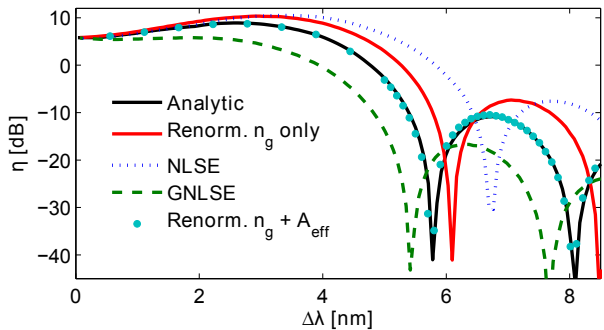}}
\caption{$\eta=P_{idler}(L)/P_{signal}(0)$ for various pump-signal detuning after a propagation of $L=100\mu m$. Black: Analytic model. Thin dashed blue: results of the NLSE. Thick dashed green: GNLSE. red: GNLSE renormalized where only the group index variations are taken into account. cyan dots: GNLSE where both slow light and effective area variations are included into the renormalization.}
\label{Extrafig} 
\end{figure}

These coefficients are computed for each pump-signal detuning $\delta\omega$; as a result we get the FWM gain curve as shown in thick black in \fig{Extrafig}. We compare now this analytic curve to the results given by the different models.
%%%%%%%%%%%%%%%%%%%%%%%%%%%%%%%%%%%%%	extra %%%%%%%%%%%%%%%%%%%%%%%%%%%%%%%%
%%%%%%%%%%%%%%%%%%%%%%%%%%%%%%%%%%%%%%%%%%%%%%%%%%%%%%%%%%%%%%%%%%%%%%%%%%%%%%%%%%%

\par First we see that the GNLSE (thick dashed green) does not appear to converge any better than the standard NLSE (light dashed blue). The NLSE does not take into account any variations of the nonlinearity, hence $\gamma_{SPM}=\gamma_{XPM}=\gamma_{FWM}$. While the NLSE tends to overestimate the FWM bandwidth, the GNLSE underestimates the FWM gain.  Thus the inclusion of self-steepening (i.e. the dispersion of the nonlinearity) does not appear as a great improvement. One must note that the GNLSE is still essential to model effect specifically related to self-steepening like the formation of a shock front or to explain the energy dependent time-advance of the nonlinear pulse \cite{Raineri_PRB13, colman:PRA14}. We see that the renormalization of the slow light enhancement factor (thick red) improves the convergence of the GNLSE: for small detuning ($\delta\lambda<2nm$) the GNLSE and the analytic model converge; for larger detuning some discrepancies still remain, but the overall error is still weaker than for the un-renormalized GNLSE or the simple NLSE. The remaining error is due to the fact that the effect of the variation of the effective modal area is neglected at first. We see that if we then renormalized the GNLSE to take as well into account that later effect (cyan dots), we obtain a very good agreement between the analytic model and the renormalized GNLSE. We present in the last section how the renormalization technique is general and can take into account variation of the modal area.
\par This demonstrates that the renormalization technique that we present here really constitutes an improvement compared to previous formulations. More generally our technique could also be seen as a generalization of \linecite{Santagiustina:10} that deals in a seamless way with the generation of multiple signal and idler orders, hence not only limited to a discrete set of a few (usually four) beams.

\subsection{Discussion}
\label{sec:Discussion}

%%% Apres : dispersive NL (shock or not shock) + XPM//SPM discussion
\par \delete{The two models are not equivalent, meaning that there are some differences in the way the physics is taken into account.}
\add{The differences between the two models (with or without the renormalization) indicate that the physics governing the nonlinear pulse is necessarily different.} Especially, one of the largest changes between \eq{GNLSE} and \eq{NLSE:Renormed} lies in the ratio of Self-Phase (SPM) to Cross-Phase Modulation (XPM) intensity. Usually the XPM is twice as many times stronger as the SPM for a given frequency. After the renormalization of the slow light variations, the photons are weighted by the quantity  $m(\omega)^2=n_g(\omega)/n_0$ and they do not have the same contribution to the nonlinear index change. The weight factor corresponds to the increase of the photonic density of states in the PhCWGs compared to the bulk material. Indeed the first derivative of the band diagram could be both interpreted as the group velocity or the density of states: photons with a higher density of states will have a higher probability to interact with the nonlinear medium. By analogy, the inclusion of the variations of the effective modal area would correspond to dividing the photon eigen contribution by their effective volume. Thus the enhancement of the nonlinearity due to the slow light and the tight confinement of light can be seen as an enhancement caused by the increase of the optical density of states. The photons with a high density of states are also less influenced by the other photons with a lower density of states ($SPM<2XPM$).
\par Another interesting effect lies in the magnitude of the shock term : the stronger this term is, the faster will the pulse form a shock front. The formation of an optical shock -more precisely the presence of the Kerr shock term- might play a very important role in the generation of dispersive linear waves (DSW) that could then be generated without strong dispersion requirements, for instance the presence of ZVD point is not mandatory\cite{Biancalana:PRA14}. The intensity of the XPM and its dependence with the angular frequency plays then a predominant role. After the renormalization, the dispersive nonlinearity ($\gamma_1/\gamma_0$) in PhCWGs is only about one third of its nominal value in \eq{GNLSE}; while at the same time the weight of the photons has been strongly modified. Consequently the behavior of PhCWGs with regards to this new DSW generation scenario is different to what is found in other systems. More generally, the renormalization redefines in a non-trivial way the interaction, mediated by the material nonlinearity, between the photons. Consequently, in the laboratory reference frame (i.e. considering only the power flux and no renormalization) the PhCWG behavior would be different to what is \textit{a priori} expected.

\section{Impact of Mode area}
\label{sec:gh3}

\par \delete{Thus far, we only considered the variations of the slow light factor; and neglected the impact of variations of the modal area. The main reason is that the slow-light is responsible for most of nonlinearity variations in PhCWGs. However the changes in the effective modal area still account for about 25\% of the dispersive nonlinearity. This situation is reverted in other systems like nanowires, which exhibits very little slow light, and where the effective modal area variation is the most important.}\add{Thus far, we only presented how to include the variations of the slow light factor in the renormalization process. The main reason is that the slow-light is responsible for most of nonlinearity variations in PhCWGs. However the change in the effective modal area still accounts for about 25\% of the dispersive nonlinearity; and we have shown through the comparison with an analytic test case that it has a noticeable impact on the overall nonlinear dynamics.} Consequently we now show how the renormalization method is also able to include variations of the effective modal area.\\ By analogy with what has been done in the previous section, the renormalization procedure is applicable as well for the modal area subject that  $gh_{\omega0}$ can be decomposed as $gh_{\omega0}(\omega_1,\omega_2,\omega_3)=g(\omega0)h(\omega_1)h(\omega_2)h(\omega_3)$. If such is the case, then we define the renormalization function $m(\omega)=h(\omega)\sqrt{n_g(\omega)}$, and the propagation equation to solve becomes:

\begin{widetext}
\begin{align}
	\partial_z \Psi(z,\omega_0) &=  \imath K(\omega_0) \Psi(z,\omega_0)
	&+\imath \frac{\omega_0n_{2I}}{cA_{eff}(\omega0)}\frac{n_{g\omega_0}}{n_0} g(\omega_0)h(\omega_0)\int&\Psi(z,\omega_1)^*\Psi(z,\omega_2)\Psi(z,\omega_3) \delta(\omega_2+\omega_3-\omega_1-\omega_0)d\omega_i^3
	\label{NLSE:Renormedgh}
\end{align}
\end{widetext}

Consequently $h(\omega)$ weights the individual photons' contribution to the Kerr nonlinearity: it includes implicitly most of the dispersive nonlinearity (in PhCWGs). On the contrary $g(\omega_0)h(\omega_0)/A_{eff}(\omega_0)$ \textbf{stands for the dispersive part of the modal area}. The problem of the factor decomposition of $gh_{\omega0}$ is directly related to the question whether the nonlinearity in PhCWGs can be modelled accurately using an analytic function \cite{6123170}. It has been demonstrated that, for dispersion engineered PhCWGs like the one considered in the present paper, the nonlinearity can be fitted by a Morse type potential function with four adjustable parameters. However the decomposition of such function in factor decomposition will only gives an approximate. More generally the best way to decompose the variations of the effective modal area is still an open question. \\ In any case, it is always possible to choose a decomposition that preserves the self-phase modulation ($SPM(\omega)\propto g(\omega)h(\omega)^3$) and cross-phase modulation $XPM(\omega_{ref},\omega)\propto g(\omega)h(\omega_{ref})^2h(\omega)$. $\omega_{ref}$ is defined as the center of the frequency domain.

\begin{align}
h(\omega)=&\sqrt{ \frac{SPM(\omega)}{XPM(\omega_{ref},\omega)} }& \label{h_def}\\
g(\omega)=&\sqrt{ \frac{XPM(\omega_{ref},\omega)^3}{SPM(\omega)^3} }& \label{g_def}
\end{align}

\begin{figure}[htb]
\centerline{\includegraphics[width=9cm]{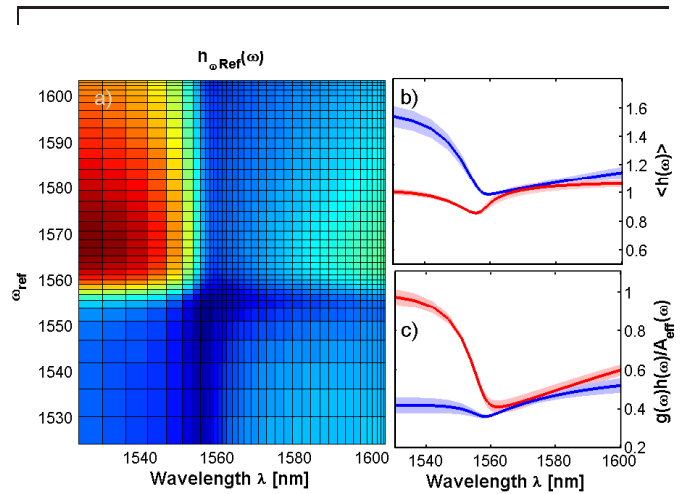}}
\caption{a) Color map: $h(\omega)$ computed for different $\omega_{ref}$ (y-axis). b) $h(\omega)$ averaged on the different $\omega_{ref}$ for the normal dispersion region (blue); and the anomalous dispersion region (red). Shaded area indicates the standard deviation of the average. c) Same as b), but showing the dispersive part of the nonlinearity $g(\omega)h(\omega)/A_{eff}(\omega)$.}
\label{fig-gh}
\end{figure}

If the decomposition is consistent -i.e. if $gh{\omega_i}$ can indeed be decomposed in factors-, $h(\omega)$ and $g(\omega)$ do not depend on the central frequency $\omega_{ref}$ that is chosen. Otherwise, only the XPM created by a pulse centered at $\omega_{ref}$, as well as the SPM for any frequency, are included correctly. Such approximation could still be sufficient if the propagation is dominated by a single strong pulse. Checking how $h(\omega_{ref})$ and $g(\omega_{ref})$ depends on $\omega_{ref}$ is essential to assess the validity of the renormalization for taking into account effects related to the effective modal area. In \fig{fig-gh}-a), we show the value of $h(\omega)$ computed according to \eq{h_def} depending on the central frequency $\omega_{ref}$ (Y-axis).

%%%%%%%%%%%%%%%%%%%%%%%%%%%%%%%%%%%%%	FIGURE	4 %%%%%%%%%%%%%%%%%%%%%%%%%%%%%%%%
%%%%%%%%%%%%%%%%%%%%%%%%%%%%%%%%%%%%%%%%%%%%%%%%%%%%%%%%%%%%%%%%%%%%%%%%%%%%%%%%%%%

We observe two main zones: one ranging $1525-1560$nm (anomalous dispersion) and the other one ranging $1560-1610$nm (ranging from the first zero GVD to the band edge). The two zones are separated by the first zero group velocity dispersion (GVD) point (cf. \fig{Parameter}-a)). Inside each zone, $g(\omega)h(\omega)$ and $h(\omega)$ do not depend on the central reference frequency as seen in \fig{fig-gh}-b,c).\\ This brings forth two major conclusions. First it appears that inside each zone, the decomposition of $gh(\omega_i)$ holds; and it is therefore possible to describe correctly the pulse evolution using the renormalization technique (of course providing that the simulation spectral domain being confined within one zone). Second, the presence of two distinct and well defined zones indicates that there is actually a change in the physics governing the photons evolution.\\ For high frequencies (small wavelengths), the photons have a nonlinear dispersive behavior corresponding to the fact that $h(\omega)g(\omega)/A_{eff}(\omega)$ varies while $h(\omega)$ is almost constant (\fig{fig-gh}-b), blue). This indicates that the photons have a quasi equi-contribution to the nonlinearity. On the contrary, for low frequencies (long wavelengths), the individual contribution of photons is more pronounced (\fig{fig-gh}-b), red) and $h(\omega)$ exhibits variations up to 60\% while $h(\omega)g(\omega)$ remains flat. Weighting the eigen photons contribution to the nonlinearity is essential. \\ In \textit{dispersion-engineered} PhCWGs, the slow light at long wavelength (low frequency) is caused by the presence of a complete Photonic Bandgap (PBG). Close to the band edge, the physics is then similar to what is found in Bragg Gratings \cite{Eggleton:99}. In contrast, the slow-light obtained at higher frequency through \textit{dispersion-engineering} has another nature and arises thanks to the complex interferences occurring inside the Bloch mode. Although the decomposition of $gh(\omega)$ into a product of functions is not mathematically exact, we see here that it could nevertheless be a useful metric to sort-out slow light \cite{Boyd:11} into categories depending on the equi-contribution to the nonlinearity that the photons have -or do not have-.
\par In the introduction, we disregarded the coupled set of equations as a possible solution for modeling systems with large variation of the nonlinear parameters. The fact that the photons are split here into two well defined domains tends to rehabilitate a posteriori such strategy. Still one must be careful on the way photons are taken into account at the limit between the two domains, especially considering that this point is precisely the zero group velocity dispersion (ZVD) point.\\ Finally we focused on SPM, XPM in the factor decomposition of $gh(\omega_i)$. \add{The photon weight $h(\omega)$ depends on this peculiar choice. Therefore a material exhibiting a different nonlinear response, like the presence of a $\chi^{(2)}$ nonlinearity or a different form for the $\chi^{(3)}$ tensor, would lead to a different $h(\omega)$.}

\bigskip
\section{Conclusion}
\label{sec:Conclusion} 

\par We have presented a new method to incorporate in an efficient way into the GNLSE the variations of the nonlinearity that exist in systems with \textbf{structured slow-light}: effects linked to variation of the slow-light enhancement factor could be taken into account through renormalization. Fluctuations of the mode effective area can be dealt with by this technique as well. This would be of importance especially for system like nanowires, where variations of the effective mode area are important. As a result this paper gives practical hints regarding the way nonlinear pulse propagation in nanostructured systems could be computed; and what could be the limitation of current models.\\ This is crucial as more studies are precisely focusing on higher order nonlinear effects and their mutual interplay \cite{Biancalana:PRA14,Husko_ScRep13,Raineri_PRB13,Colman2012,colman:PRA14}. Our model is consistent with an analytic set of equations derived to model discrete FWM events \cite{Santagiustina:10}; and could be considered as a generalization of that article. Besides, it is worth to note that our technique does not increase the computational burden compared to the resolution of a standard GNLSE.

\par This study was also the occasion for more fundamental considerations. In particular we found relevant not to compute directly the propagation of the power flux but to weight first the photon contribution by their optical density of states. Although the modal area depends on the considered nonlinear effect and its associated tensor, we found two classes of slow light in \textit{dispersion-engineered} PhCWGs: engineered slow light with a normal nonlinear dispersive behavior, and a region close to the photonic band gap where the weight factor of the photons contributes greatly. Such study might change how we perceive and understand slow-light effects which are in fact more related to \textit{high density of states light} effect.

\par \delete{The renormalized equation does not hold any new physics in itself, but the fact that the propagation is solved in a frame different from the laboratory frame (i.e. power flux) leads to a different nonlinear dynamics. Especially the balance between SPM and XPM (and the symmetry of the XPM) is altered and some photons appear more immune to perturbation, or prone to perturb other photons, depending on their relative weight $m(\omega)$. That latter aspect requires further investigations. Still PhCWGs appear already a new very peculiar systems for nonlinear optics from a fundamental, and not only applied, point of view.}\add{Usually, the improvement of the models which deal with nonlinear pulse propagation comes along with the addition of extra operators in order to describe the new effects. Here the renormalized equation keeps the formulation with no extra terms added. Indeed, the new phenomena that are observed lie inside the renormalization function $m(\omega)$, not in the GNLSE itself. Namely the unbalance between SPM and XPM could be interpreted as inertial forces that appear because of the non-trivial relationship between the laboratory referential and the PhCWGs one, where some photons appear more immune to perturbation or prone to perturb others. Within this new reference frame, all the semi-analytic method/models developed so far, like the momentum method, remain valid.}

\par Finally the present discussion only focused on PhCWGs, a system where the nonlinear variations are extreme. Part of our conclusions would anyway also apply to other nano-structured systems like nanowires where the relatively weaker nonlinear variations must be seen in regards to the very large optical bandwidth these systems support. Especially we have shown that variations of the effective area -which are dominant over slow-light in nanowires- can be taken into account by the renormalization method; and that this method can describe with accuracy both self-phase and cross phase modulation effects.

\bigskip
\noindent \textbf{Acknowledgements}\\ This work has been supported through the Villum Fonden funded centre of excellence NATEC. 
%The author thanks Chad Husko, Stefano Trillo, John E. Sipe and Aloyse Degiron for useful discussions.

%\pagebreak
\bibliography{Bibliography}{}

\end{document}